\documentclass{ar2e}
\begin{document}
%\input epsf.def   %<-If you need EPS figures to be
                  %  called in {figure} environment.
\input psfig.sty

\bibliographystyle{unsrt} %for BibTeX - sorted numerical labels by
			  %order of first citation.

%\arraycolsep1.5pt

% A useful Journal macro
\def\Journal#1#2#3#4{{#1} {\bf #2}, #3 (#4)}

% Some useful journal names
\def\NCA{\it Nuovo Cimento}
\def\NIM{\it Nucl. Instrum. Methods}
\def\NIMA{{\it Nucl. Instrum. Methods} A}
\def\NPB{{\it Nucl. Phys.} B}
\def\PLB{{\it Phys. Lett.}  B}
\def\PRL{\it Phys. Rev. Lett.}
\def\PRD{{\it Phys. Rev.} D}
\def\ZPC{{\it Z. Phys.} C}

% Some other macros used in the sample text
\def\st{\scriptstyle}
\def\sst{\scriptscriptstyle}
\def\mco{\multicolumn}
\def\epp{\epsilon^{\prime}}
\def\vep{\varepsilon}
\def\ra{\rightarrow}
\def\ppg{\pi^+\pi^-\gamma}
\def\vp{{\bf p}}
\def\ko{K^0}
\def\kb{\bar{K^0}}
\def\al{\alpha}
\def\ab{\bar{\alpha}}
\def\be{\begin{equation}}
\def\ee{\end{equation}}
\def\bea{\begin{eqnarray}}
\def\eea{\end{eqnarray}}
\def\CPbar{\hbox{{\rm CP}\hskip-1.80em{/}}}%temp replacemt due to no font

\def\beqn{\begin{equation}}
\def\eeqn{\end{equation}}
\def\beqna{\begin{eqnarray}}
\def\eeqna{\end{eqnarray}}

%%%%%%%%%%%%%%%%%%%%%%%%%%%%%%%%%%%%%%%%%%%%%%%%%%%%%%%%%%%%%%%%%%%%%%%%
%%BEGINNING OF TEXT                           
%%%%%%%%%%%%%%%%%%%%%%%%%%%%%%%%%%%%%%%%%%%%%%%%%%%%%%%%%%%%%%%%%%%%%%%%

\jname{Annu. Rev. Nucl. Part. Sci.}
\jyear{2001}
\jvol{51}
\ARinfo{1056-8700/97/0610-00}

\title{Parity-Violating Electron Scattering and Nucleon Structure}

\markboth{R. D. McKeown}{Parity-Violating Electron Scattering and Nucleon Structure}

\author{D. H. Beck
%\thanks{Representing the SAMPLE collaboration\cite{collab}}
\affiliation{Department of Physics, 
University of Illinois at Urbana-Champaign, \\
Urbana, IL, 61801, USA, \\
E-Mail: dhbeck@uiuc.edu}
R. D. McKeown
%\thanks{Representing the SAMPLE collaboration\cite{collab}}
\affiliation{W. K. Kellogg Radiation Laboratory, 
California Institute of Technology, \\
Pasadena, CA 91125, USA, \\
E-Mail: bmck@krl.caltech.edu}}%\vskip 0.5 truein
%\centerline{Representing the SAMPLE Collaboration${}^1$}
%\bigskip


\begin{abstract}
The measurement of parity violation in the helicity
dependence of electron-nucleon scattering 
provides unique information  about the
basic quark structure of the nucleons.
In this review, the general formalism of parity-violating
electron scattering is presented, with emphasis on elastic 
electron-nucleon scattering. The physics issues addressed by
such experiments is discussed, and the major goals
of the presently envisioned experimental program are identified.
%General aspects of the experimental technique are reviewed and 
A summary of results from a recent series of experiments is
presented and the future prospects of this program are also discussed.
\end{abstract}

\maketitle

\section{INTRODUCTION}
%{Introduction}

The study of the parity-nonconserving force between electrons and
quarks has been of fundamental importance in exploring the nature
of the neutral weak interaction mediated by
the $Z$ boson. The 
seminal experiment performed at SLAC in 1976 \cite{prescott}
not only confirmed the Lorentz structure of
the neutral weak interaction 
(along with atomic parity violation 
experiments \cite{atomic}) 
but also introduced a powerful new 
experimental technique: the measurement of helicity-dependence in 
electron scattering.

During the last quarter century
the standard electroweak theory has been established with phenomenal
quantitative success, and
now provides a firm basis for the use of the weak
interaction as a precision probe of nucleon structure.
In particular, the parity-violating interaction of electrons with nucleons
can provide unique and novel information on the weak structure
of nucleons and their associated quark structure \cite{bmck89,beck89}. 
This includes sensitivity
to strange quark-antiquark effects \cite{kaplan} and to higher order 
effects \cite{musolf90} such as the anapole moment. 
In this review, the development
of this subject is presented, including the theoretical framework 
and the associated 
experimental program.

The importance and historical interest in the $\bar s s$ contributions to 
nucleon structure are briefly reviewed in the next section. The sensitivity
of the neutral weak form factors 
determined in parity-violating
electron scattering to the sea quark structure of
the nucleons discussed in Sections 3 and 4 below. In addition,
Section 5 treats the role of
the axial form factor of the nucleon as measured in
parity-violating $e^-$-$N$ elastic scattering, which provides
access to higher order processes such as the anapole form
factor and electroweak radiative corrections. 
Thus the study of parity-violating electron scattering offers an 
advantageous method to access this novel aspect of weak nucleon structure
and provides a sensitive testing ground for calculations of electroweak
corrections beyond leading order in perturbation theory.

There are also corresponding advances in the
experimental methods employed to study the small parity-violating observables,
and the present 
and future experimental program to explore this subject is reviewed
in Section 6.

\section{STRANGE QUARKS IN THE NUCLEON}
%{Strange quarks in the nucleon}

For historical reasons, the role of strange
quark-antiquark pairs in nucleon structure had been ignored for many
years.
Traditional
constituent quark models were rather successful in
treating the nucleon using only up and down
quarks. But it is important to remember that, in this approximation, only
the degrees of freedom
associated with valence quark quantum numbers
are active; the effects of sea quarks are generally
considered to be ``frozen'' as inert aspects of the effective
degrees of freedom. The constituent quarks actually do contain
internal structure associated with gluons and sea quarks (such as $\bar s s$), 
so even in these simple models of the nucleon the contribution
of $\bar s s$ pairs is potentially significant and therefore
interesting.

In the following discussion, a variety of evidence for the presence of
strange quark-antiquark pairs in the nucleon is reviewed in order
to provide the appropriate perspective for consideration of the
$\bar s s$ content of the neutral weak form factors, to be discussed below in
Section 4.

\subsection{Deep Inelastic Neutrino Scattering}

\begin{figure}
\centerline{\psfig{figure=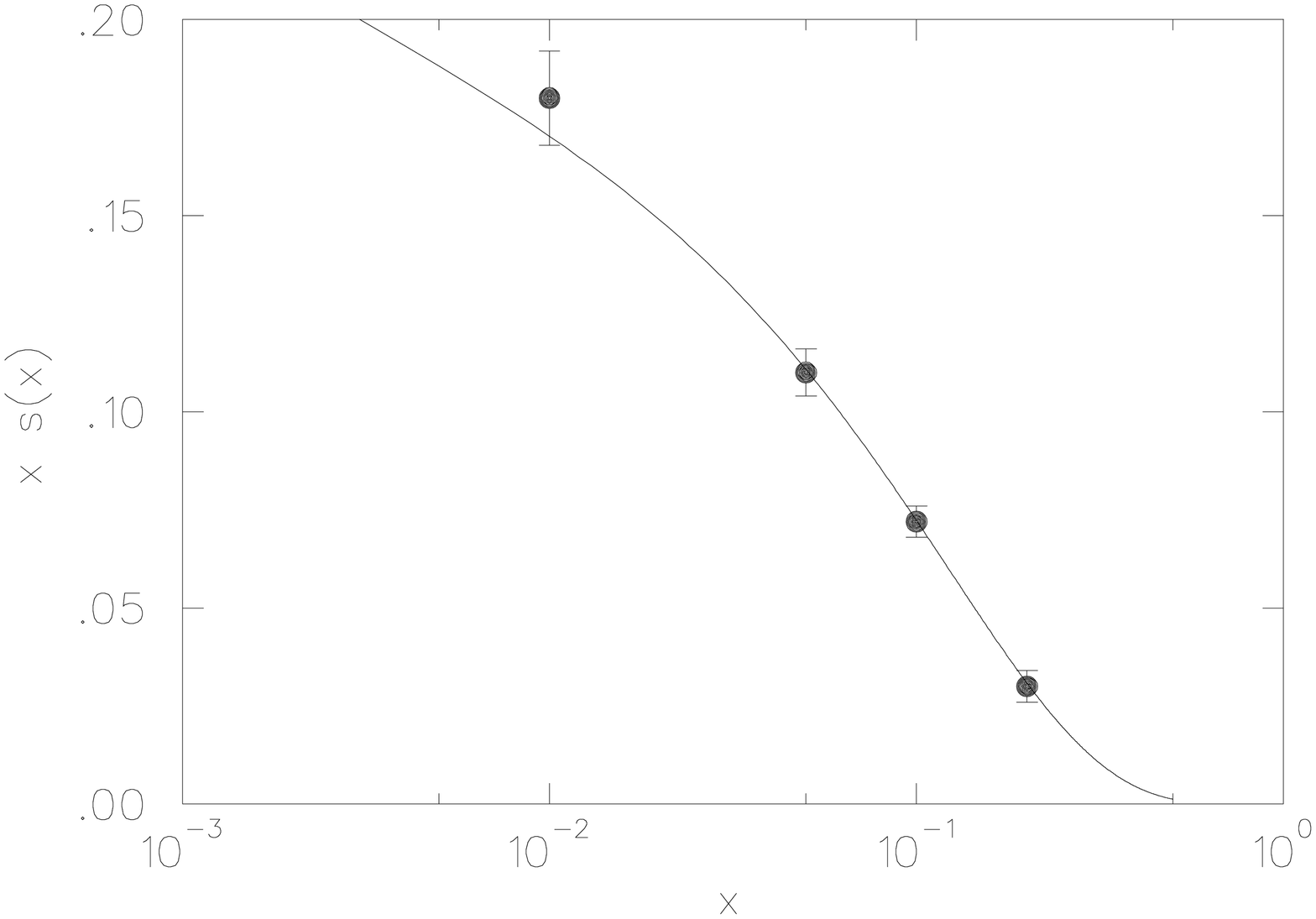,angle=90,height=3.0in}}
\caption {Measured values of $x \cdot s(x)$ at renormalization
scale $\mu^2=4$ (GeV)${}^2$
from a next to leading order analysis of
deep inelastic neutrino scattering \cite{bazarko}. There is no
significant difference observed between $s(x)$ and $\bar s (x)$ so
this analysis has assumed they are equal.}
\label{fig:sx}
\end{figure}

The most direct method
of detecting the presence of quarks in nucleons is to employ deep inelastic
lepton scattering. The quark structure of the nucleon is
described through the use of structure functions which can be
determined by measurements of the deep inelastic scattering
cross sections. These structure functions depend on the
Bjorken scaling variable $x$ which is interpreted as the fraction
of nucleon meomentum carried by the struck quark in the infinite
momentum frame. Thus, the nucleon quark structure is expressed
in terms of the individual quark structure functions $u(x)$, $d(x)$,
$\bar u(x)$, $\bar d(x)$, $s(x)$, $\bar s(x)$, etc. 
Elucidation of the flavor structure is facilitated
by using charged-current neutrino and antineutrino interactions. 
Neutrinos interact with $d$ and $s$ quarks by raising their
charge and producing a negative lepton (e.g. $ \nu_\mu + d \rightarrow
\mu^- + u$ or $ \nu_\mu + s \rightarrow \mu^- + c$). The charmed quarks
produced by the $s$ quarks then decay semileptonically yielding
$\mu^+$'s, and so one observes $\mu^- \, \mu^+$ pairs from $\nu_\mu$
interactions with $s$ quarks. Similarly, antineutrinos will produce
$\mu^+ \, \mu^-$ pairs from $\bar s$ quarks. In this way, measurements
of $s(x)$ and $\bar s(x)$ have been performed in deep inelastic
neutrino and antineutrino experiments \cite{bazarko}. The results,
shown in Figure~\ref{fig:sx}, indicate that $s(x)$ and $\bar s(x)$ are
significant 
at low $x<0.1$ and the $s$ and $\bar s$
each carry about 2\% of the nucleon momentum.

Although deep inelastic scattering is useful as a direct probe for the
presence of quarks, it is not possible at present to connect these
observations with the common
static properties of the nucleon such as the mass, spin, or magnetic
moment. Clearly, experimental determination of $\bar s s$
contributions to  these quantities 
would be of great interest in our attempt to understand these 
simple basic properties of the nucleon.

\subsection{Strangeness And The $\pi$-$N$ Sigma Term}

The strange quark contribution to the nucleon mass can be addressed  
by studying
the ``sigma term'' in $\pi$-nucleon scattering \cite{jaffe87,gasser}. 
One first  
obtains
the value of the isospin even $\pi$N scattering amplitude (from  
experiment)
extrapolated to the pole at $q^2=2m_\pi^2$ (the ``Cheng-Dashen'' point), 
$\Sigma_{\pi N}$. The most recent analyses \cite{gasser} indicate 
that a value of $ \Sigma_{\pi N}(2m_\pi^2) \sim 60 $ MeV is obtained
from the experimental data. This value is then extrapolated to
$q^2=0$ to give  $ \Sigma_{\pi N}(0) \sim 45 $ MeV.
Then 
one may utilize hyperon mass relations (assuming the strange quarks
contribute linearly in the strange quark mass $m_s$) to  
arrive at a prediction for the quantity 
\begin{equation}
\sigma \equiv {1 \over {2M_p}} \langle p | \hat m (\bar u u + \bar d
d) | p \rangle \simeq 25 \> {\rm {MeV}} \> ,  
\end{equation}
where $\hat m \equiv (m_u + m_d)/2$. 
SU(3) corrections to $\sigma$ are expected to be about 10 MeV,
yielding a predicted value of 
\begin{equation}
\sigma_0 \simeq 35 \> {\rm MeV} \> .
\end{equation}
In the absence of a contribution from $\bar s s$ pairs in the nucleon,
one would expect $\Sigma_{\pi N}(0) = \sigma_0$. 
The recent detailed  
analysis by Gasser {\it et al.} \cite{gasser} examines the 
various corrections and
extrapolations
mentioned above. These authors conclude that the difference between
$\Sigma_{\pi N}(0)$ and $\sigma_0$ implies a finite
contribution
of strange quark-antiquark pairs to the nucleon mass
\begin{equation}
m_s \langle p | \bar s s | p \rangle \sim 130 \, {\rm MeV} \>.
\end{equation}
However, one should treat this result with caution for several
reasons. There are questions associated with the accuracy of the 
the extrapolation of the $\pi$N
amplitudes into the unphysical region, 
the precision and consistency of the $\pi N$ data, 
and the effect of SU(3) symmetry
breaking in the hyperon mass relations. 
Indeed more recent studies \cite{liu,thomas} of this problem using some input from
lattice QCD, but using different techniques, indicate that the strangeness contribution to the nucleon
mass may be about twice as large as that quoted above~\cite{liu}, or 
nearly zero~\cite{thomas}.  Further, recent 
analyses of new experimental work~\cite{olsson,pavan} suggest $\Sigma_{\pi N}$
may be significantly larger than previously thought.
All of these concerns lead to
a reduced confidence in the quoted result, and the uncertainty in the
quoted value of
the matrix element $\langle p | \bar s s | p \rangle$ is probably of
order 100\%.

\subsection{Strangeness And Nucleon Spin}
\label{sec:nucleonSpin}

The flavor structure of the nucleon spin can be addressed by studying  
spin-dependent deep-inelastic lepton scattering. An extensive series
of experiments using electron and muon scattering over the last decade
have produced an impressive set of data
 \cite{dis}. The spin dependent quark structure is described
by spin dependent structure functions $u^+(x)$, $u^-(x)$, $d^+(x)$,
$d^-(x)$, $s^+(x)$, and $s^-(x)$ where $+$ (or $-$) refers to the quark
spin being parallel (or antiparallel) to the nucleon spin. One usually
defines the integrated differences 
\begin{equation}
\Delta u \equiv \int_0^1 \lbrack  u^+(x) - u^-(x)\rbrack \, dx 
\end{equation}
and similarly for $\Delta d$ and $\Delta s$. 
One can experimentally determine the
spin dependent structure function of the nucleon $g_1(x)$
and then compute the first moment 
\begin{equation}
\Gamma_1 \equiv \int^1_0 g_1(x) \,  dx \> .
\end{equation} 
For the proton, this quantity is related to $\Delta u$, $\Delta d$,
and $\Delta s$ by
\begin{equation}
\Gamma_1^p = {1 \over 2} \left( 
{4 \over 9} \Delta u + {1 \over 9} \Delta d + {1 \over 9}
\Delta s \> \right) .
\end{equation}
All of these expressions are modified at finite $Q^2$ by QCD
corrections that introduce mild but significant $Q^2$ dependence
 \cite{disfit}.

Having obtained $\Gamma_1^{p,n}$ from experiment, one can combine this
information with the isovector axial matrix element (known from neutron beta
decay \cite{pdg})
\begin{equation}
G_A(Q^2=0) = \Delta u - \Delta d = F+D = 1.2601 \pm 0.0025
\end{equation}
and the octet combination (from hyperon beta decays \cite{jaffe90})
\begin{eqnarray}
a_8 &=& 
%{1 \over {2 \sqrt 3}} 
(\Delta u + \Delta d -2 \Delta s) = 3F-D \\
& =& -0.60 \pm 0.12 \> \nonumber
\end{eqnarray}
to obtain values for the individual flavor components $\Delta u$,
$\Delta d$, and $\Delta s$.
The most recent analyses of the experimental
results \cite{disfit} indicate that the fraction of the nucleon spin carried by
quark spins is roughly 
\begin{equation}
\Delta u + \Delta d + \Delta s  = 0.20 \pm 0.10  \> .
\end{equation}
That is, only 20\% of the nucleon's spin is carried by the quark
spins.

The actual contribution of strange
quarks is more difficult to extract due to substantial sensitivity 
to SU(3) breaking effects. However, the recent analyses
 \cite{dis,disfit} all 
tend to favor a result that is in the range
\begin{equation}
\Delta s \simeq -0.1 \pm 0.1 \> .
\end{equation}
While this seems rather small, it may actually be quite significant compared
to the total spin carried by the quark spins in Eqn. (9).
Nevertheless, concerns
associated with SU(3) breaking, the extrapolation of the data to
$x=0$ to form the integrals in Eqn. (5), and the uncertainties arising
from the $Q^2$ evolution of the
structure functions all contribute to reduced confidence in the
extraction of $\Delta s$ using this method.

It has been suggested that the strange spin-dependent
structure function $\Delta s(x)$ could be determined in measurements
of semi-inclusive kaon production in 
spin-dependent deep inelastic scattering \cite{milner}.
Although the problem of extrapolation to $x=0$ will remain, this
information would be very helpful in further
constraining $\Delta s$. Of course, additional uncertainties associated
with factorization of the fragmentation process may be encountered
in the interpretation of the semi-inclusive asymmetries.
The HERMES experiment at DESY will acquire data on this reaction
in the near future \cite{hermes}.

It is also possible to obtain information on $\Delta s$ from elastic
neutrino nucleon scattering. As discussed in Sections 3 and 5, 
the quantity $\Delta s$ contributes to 
the neutral axial 
vector coupling of the nucleon at $Q^2$=0. Recent efforts to analyze
existing $\nu_\mu$-$p$ data \cite{garvey} indicate that the uncertainties
are too large to make a meaningful statement. However, measurements
at very low $Q^2 \le 0.1$ (GeV/c) would be very sensitive
to the axial form factor of interest.
Such measurements are technically very difficult, but it has been
suggested \cite{garvey92} that one should 
measure the ratio of $\nu$-p to $\nu$-n cross sections 
in
quasielastic scattering from a $Z=N$ nucleus like Carbon.
In the $Q^2=0$
limit the ratio for free nucleons is approximately
\begin{equation}
{{\sigma_p} \over {\sigma_n}} \simeq 1 - 2{{\Delta s} \over {G_A}}\> .   
\end{equation}
It has been shown that the nuclear corrections to this ratio 
in Carbon are quite small \cite{garvey92}.  The feasibility of such a 
measurement will be investigated in the mini-Boone experiment at Fermilab~
\cite{tayloe}.

\section{PARITY VIOLATION IN ELECTRON SCATTERING:  THEORETICAL FORMALISM}
%{Parity Violation in Electron Scattering: Theoretical Formalism}

\begin{figure}
%\centerline{\epsfysize=5.0in \epsfbox{gamma_z.eps}}
\centerline{\psfig{figure=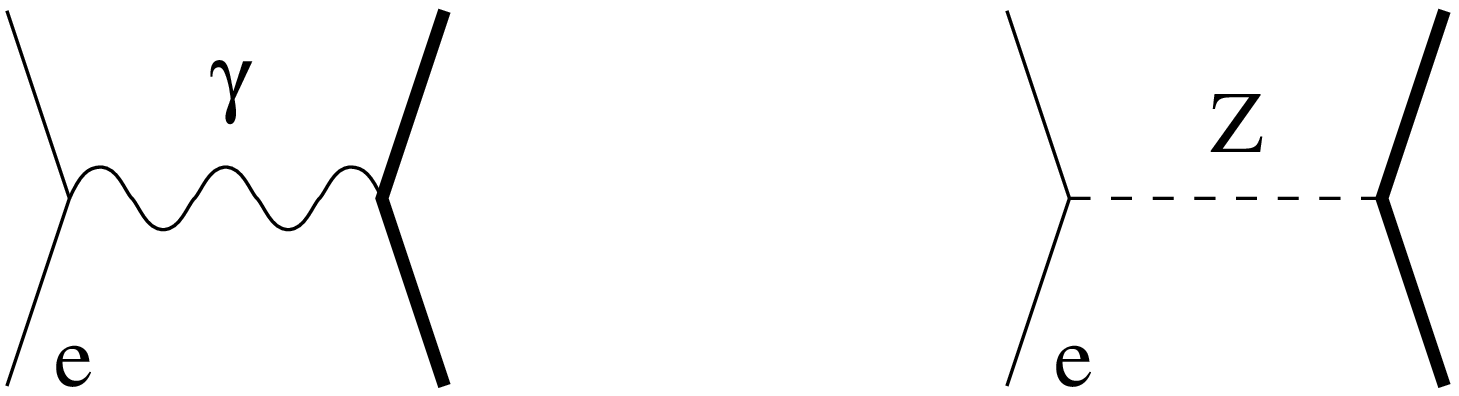,angle=270,width=5.0in}}
\caption {The amplitudes relevant to parity-violating electron
scattering. The dominant parity-violating effects arise from
the interference of these two  
amplitudes.}
\label{fig:amplitudes}
\end{figure}

The essential aspects of the parity-violating interaction between 
electrons and other electrically charged
objects (generally hadrons composed of quarks
or another electron) can be seen in Figure~\ref{fig:amplitudes}. The 
dominant amplitude arises from the electromagnetic interaction
(photon exchange) whereas the parity-violating neutral weak interaction
corresponding to $Z$ boson exchange generates a small amplitude 
that is detectable via quantum interference. Due to the parity-violating
nature of the weak interaction, these interference effects
imply the existence of small pseudoscalar observables in 
electron scattering experiments. 

In the standard electroweak model, the Lorentz structure of
the neutral weak interaction is a linear combination of vector and
axial vector couplings to the $Z$ boson. Aside from the overall
normalization governed by the fundamental electric charge $e$
and the $Z$ boson mass $M_Z$, the electron and various flavors of quarks
have vector and axial vector coupling strengths that depend 
upon the weak mixing angle
$\theta_W$ as illustrated in Table~\ref{tab:couplings}. Here the relative 
strengths of the couplings (in lowest order $\gamma$- and
$Z$- exchange) of the
basic objects of interest in this discussion are listed;
the vector interactions are denoted by charges $q$ 
($q^\gamma$ for $\gamma$-exchange and $q^Z$ for $Z$-exchange) and the axial
charges are referred to by $a^Z$ in this table.

\begin{table}
\caption{Electroweak couplings of charged fundamental particles}
\label{tab:couplings}
\vspace{0.2cm}
\begin{center}
\footnotesize
\begin{tabular}{lrcr}
\hline
\noalign{\vskip3pt}
{ }&$q^\gamma$ &$q^Z $&$a^Z$\\
\noalign{\hrule}
\noalign{\vskip6pt}
e&$\ {-1}$&
$\ -({1}-{4}\sin^2\theta_W)$&$+1$\cr
u&$\ {2\over3}$&
$\ {1}-{8\over3}\sin^2\theta_W$&$-1$\cr
d&$-{1\over3}$&$-1+{4\over3}\sin^2\theta_W$&$+1$\cr
s&$-{1\over3}$&$-1+{4\over3}\sin^2\theta_W$&$+1$\cr
\noalign{\vskip6pt}
\noalign{\hrule}
%\hline
\end{tabular}
\end{center}
\end{table}

One very important feature evident in Table~\ref{tab:couplings} is
that the weak vector couplings $q^Z$are dependent on the 
weak mixing
angle $\theta_W$. In the standard electroweak model, this parameter
is related (in lowest order)
to the neutral ($Z$) and charged ($W$) boson masses by
the relation
\begin{equation}
\cos \theta_W = {{ M_W} \over {M_Z}}  \> .
\end{equation}
Measurements of the $Z$ and $W$ masses, along with the
fine structure constant $\alpha$ and 
the charged weak coupling constant
\begin{equation}
G_F= {{e^2} \over {4 \sqrt{2} M_W^2 \sin^2 \theta_W}}
\end{equation}
enable a precise determination of the weak mixing angle \cite{pdg}
\begin{equation}
\sin^2 \theta_W = 0.23117 \pm 0.00016 \> .
\end{equation}
It should be noted that the value of this constant is dependent upon the
renormalization scheme, and we quote the result for the widely utilized
modified minimal subtraction
($\bar {\rm MS}$) \cite{pdg} scheme.

The couplings in Table~\ref{tab:couplings} also imply important features of
the neutral weak interactions of the composite hadrons
comprising everyday matter, protons and neutrons.
In experiments at very low energies where nucleons remain intact,
the relevant lowest order
elastic couplings to the observed physical
particles (electrons and nucleons) are as displayed in Table~\ref{tab:charges}.
Note that the separate values for magnetic ($\mu$) and charge ($q$)
interactions are listed and the magnetic moments are in units
of $e \hbar / 2 M c$ where $M$ is the particle mass. 
(The magnetic moments $\mu^\gamma$ and
weak magnetic moments $\mu^Z$ for the nucleons are significantly
altered from the Dirac values due to the internal structure of the
nucleons.) 
The neutral weak vector couplings are evaluated with the value 
$\sin^2 \theta_W = 0.23117$
 \cite{pdg}, and the contributions of strange quarks to the neutral
weak couplings $\mu^Z$ and $a^Z$  are indicated as $\mu_s$ and $\Delta s$,
respectively.
The quantities listed in Table~\ref{tab:charges} are the important 
couplings relevant for low-energy elastic lepton-nucleon
scattering experiments as well as for atomic physics experiments.

\begin{table}
\caption{Electroweak couplings of physical particles}
\label{tab:charges}
\begin{center}
\footnotesize
\begin{tabular}{lrcccc}
\hline
\noalign{\vskip3pt}
%\noalign{\hrule}
{}&$q^\gamma$ &$\mu^\gamma$& $q^Z$&$ \mu^Z$ &$a^Z$\\
\noalign{\hrule}
\noalign{\vskip6pt}
$e$&${-1}$&-1&$-0.074$&$-0.074$&$+1$\cr
$p$&$+1$&2.79& $0.074$&$ 2.08-\mu_s$&$-1.26+\Delta s$\cr
$n$&$0$&-1.91&-1&$ -2.92-\mu_s$&$+1.26+\Delta s$\cr
\noalign{\vskip6pt}
\noalign{\hrule}
%\hline
\end{tabular}
\end{center}
\end{table}

In considering parity-violating interactions of electrons one should
note that the neutral weak vector coupling of the electron
is suppressed due to the fact that $1 - 4 \sin^2 \theta_W \simeq 0.074 \ll 1$.
Therefore, the axial vector coupling of the electron is
generally dominant in such parity violation experiments.

The axial vector couplings to the light quarks listed in
Table~\ref{tab:couplings} imply a definite prediction regarding
the isospin behavior of the weak axial interaction of nucleons
displayed in Table~\ref{tab:charges}. 
In the absence of
strange quarks, it is evident from Table~\ref{tab:couplings} that the axial coupling 
is a pure isovector; in which case one would expect the 
neutral axial vector couplings of the proton and neutron to be exactly
equal (except for a sign change). 
Thus we see that in Table~\ref{tab:charges} the 
weak axial coupling to the nucleons is pure isovector except for
the contribution of strange quarks, indicated by the quantity $\Delta s$.
(This quantity is directly related
to sum rules in spin-dependent deep-inelastic electron (or muon) scattering 
discussed in Section~\ref{sec:nucleonSpin} above.)
Any finite signal associated with an {\it isoscalar}
axial coupling to the nucleons must (in lowest order) be associated
with the presence of strange quarks. 
For the case of parity-violating observables in $e^-$-$N$ interactions,
the contribution of the nucleon's axial current is multiplied by the
electron's weak charge $q^Z = -(1 - 4 \sin^2 \theta_W)$, and is therefore
suppressed.
(However, the axial current of the nucleon is significant in
elastic neutrino-nucleon scattering experiments, as
discussed later in more detail.)

The vector couplings to the quarks are also of 
great utility in the study of the flavor structure of the nucleon.
If one ignores the effects of strange quarks
the charge and magnetic interactions for the proton and neutron
are essentially reversed for $Z$ exchange relative to $\gamma$ exchange.
(This is due to the structure contained in Table~\ref{tab:couplings} coupled with the fact that
$\sin^2 \theta_W   \simeq 1/4$.) 
That is, the charge coupling to the proton is small relative to that of the
neutron and the roles of the magnetic moments are reversed 
($\mu_p^Z \sim 2 \sim - \mu_n$ and $ \mu_n^Z \sim -3 \sim - \mu_p$).
Also, one can clearly see effects of strange quark-antiquark 
pairs in the vector
nucleon couplings,
particularly the magnetic
moments. Although the absence of a net strangeness in the nucleon
implies that the strange $\bar q q$ pairs
contribute zero net electric charge (i.e., $q_s = 0$), 
there is no such prohibition for the magnetic moment and one should
expect a contribution $\mu_s$ from the strange quarks. 

Indeed, the
possible contribution of $\bar s s$ pairs to the nucleon's magnetism
is a subject of substantial current interest and 
represents the
continuation of a long-standing historical tradition
of using the magnetic structure of the nucleon as a clue to its
internal structure. (We note that 
in 1933 Frisch and Stern \cite{stern} reported the first measurement
of the magnetic moment of the proton, the earliest
experimental evidence for nucleon substructure.)
The
neutral weak magnetic moment $\mu^Z$ is just as fundamental
a property of the nucleon as the usual electromagnetic $\mu$, and it
can provide crucial new information
on the static magnetic properties of the nucleon.
Indeed, as discussed by Kaplan and Manohar
 \cite{kaplan}, the study of 
neutral weak
matrix elements (such as the neutral weak magnetic moment)
can be used to determine the strange quark-antiquark
($\bar s s$) matrix elements. Furthermore, as detailed 
in the following section, one
can perform a complete decomposition of these observables into
contributions from the three relevant quark flavors (up, down, and
strange) \cite{beck89}.

As discussed above, the presence of parity-violating observables in electron
scattering experiments is associated (to lowest order) with the
interference of $\gamma$ and $Z$ exchange amplitudes (as indicated
in Figure~\ref{fig:amplitudes}). The scattering cross  
section
will generally consist of a helicity independent piece (dominated by the 
squared electromagnetic
amplitude) and a term that depends on the electron helicity which  
violates
parity (due to the interference of electromagnetic and neutral  
weak amplitudes). 
The parity-violating weak amplitude
must contain a product of vector ($v^Z$) and an axial vector ($a^Z$) 
couplings. 
One usually quotes the ratio of helicity dependent to  
helicity
independent cross sections, or the parity-violating asymmetry:
\begin{equation}
A = {d\sigma_R - d\sigma_L \over d\sigma_R + d\sigma_L} 
\end{equation}
where $\sigma_R$ and $\sigma_L$ are the cross sections for right- and  
left-handed electrons, respectively. This quantity will generally
be proportional to a product of neutral weak couplings 
$v^Z \cdot a^Z$ that contains the physics of interest. Thus,
measurement of the helicity dependence in elastic electron-proton
scattering can be used to study the
neutral weak vector form factors of the nucleon
 \cite{bmck89,beck89}. 

 The parity-violating asymmetry for 
elastic electron-proton scattering is given by
the following expression \cite{musolf94a}:
\begin{eqnarray}
A &=& \left[- G_F Q^2 \over 4 \sqrt{2} \pi \alpha \right]
{{
\varepsilon G^{\gamma}_{{E}} G^{Z}_
{{E}} + \tau G^{\gamma}_{{M}}
G^{Z}_{{M}} - (1-4 \sin^2 \theta_W ) 
\varepsilon^{\prime} G^{\gamma}_{{M}} G^{e}_{A}}   \over
{\varepsilon (G^\gamma_
{{E}})^2 + \tau (G_{{M}})^2}} \\
 &\equiv & -{G_FQ^2\over 4 \sqrt{2} \pi\alpha}\times {{\cal N}\over 
    {\cal D}}\>
\end{eqnarray}
where
\begin{eqnarray}
\tau & = & {{Q^2} \over {4 M_N^2}} \nonumber \\
\varepsilon & = & {1 \over 1 + 2(1 + \tau)\tan^2{\theta \over 2}} \nonumber \\
\varepsilon^{\prime} & = & \sqrt{\tau (1+\tau) (1- \varepsilon^2)}
\end{eqnarray}
are kinematic quantities, $Q^2>0$ is the four-momentum transfer, and
$\theta$ is the laboratory electron scattering angle. 

The quantities $G_E^\gamma$, $G_M^\gamma$, $G_E^Z$, and $G_M^Z$ are 
the vector form factors of the nucleon associated with $\gamma$-
and $Z$-exchange.  As in the case above,
the electromagnetic and weak form factors are 
(in lowest order) related
via the flavor dependence of the fundamental couplings in 
Table~\ref{tab:couplings}.
The flavor structure of these form factors and
the radiative corrections are considered below in Section 4.

The neutral weak $N$-$Z$ interaction also involves an axial vector
coupling $G_A^e$ in the third term of the numerator in Eqn.(16).
The tree-level $Z$-exchange process is responsible for the $1-4 \sin^2 
\theta_W$ factor that appears in this expression and, as noted in 
 \cite{musolf90,musolf94a},
higher order processes can contribute significantly. These include
anapole effects and other electroweak radiative corrections as discussed
in Sections 5.2 and 5.3.

It will also be useful to consider parity-violating quasielastic
scattering from nuclear targets, particularly deuterium. We will see 
later that this will provide useful information on the (somewhat uncertain)
axial vector form factor contributions. 
For a nucleus with $Z$ protons and $N$ neutrons the asymmetry
can be written in the simple form (ignoring final state interactions
and other nuclear corrections):
\begin{equation}
A_{\rm nuc} = -{G_FQ^2\over 4 \sqrt{2}\pi\alpha}\times 
   {N{\cal N}_n + Z{\cal N}_p \over
    N{\cal D}_n + Z{\cal D}_p}
\end{equation}
where ${\cal N}_p$ (${\cal N}_n$) is the numerator expression and
${\cal D}_p$ (${\cal D}_n$) the denominator (from Eqns. 16 and 17) 
for the proton (neutron), respectively.

\section{STRANGENESS AND THE NEUTRAL WEAK VECTOR FORM FACTORS}
%{Strangeness and the neutral weak form factors}

The formalism associated
with contributions of strange quark-antiquark pairs to the charge
and magnetization distributions of nucleons and nuclei is presented in
this section.
We begin with a general discussion of the quark flavor structure of the
electromagnetic and neutral weak currents
of these objects
(in lowest order). 

The standard electroweak model
couplings to the up, down, and strange quarks were listed in 
Table~\ref{tab:couplings};
these imply that
the electromagnetic current operator has the simple form
\begin{equation}
\hat V^\mu_\gamma = {2 \over 3} \bar u \gamma^\mu u
			- {1 \over 3} \bar d \gamma^\mu d
				 - {1 \over 3} \bar s \gamma^\mu s \>.
\end{equation}
Also from Table~\ref{tab:couplings},
the neutral weak vector current operator is given by an 
analogous expression 
\begin{equation}
\hat
V^\mu_Z = (1 - {8 \over 3} \sin^2 \theta_W) \bar u \gamma^\mu u 
	+(- 1 + {4 \over 3} \sin^2 \theta_W)  \bar d \gamma^\mu d
	+(- 1 + {4 \over 3} \sin^2 \theta_W) \bar s \gamma^\mu s  \>.
\end{equation}
Here the coefficents depend on the weak mixing angle.
The flavor structure contained in these expressions forms the basis for
a program to measure the flavor composition of the vector form factors. 
The measurements involve matrix elements of these operators (the form
factors) which will reflect the underlying flavor dependence of these
operators.

\subsection{Nucleon Vector Form Factors}

The electromagnetic form factors of the nucleon 
arise from matrix elements of 
the EM current operator
\begin{equation}
 <N| \hat V^\mu_\gamma  | N> \, \equiv  \, \bar u_N \bigg[ F_1^\gamma(q^2) 
\gamma^\mu
+  {i \over {2M_N}}  F_2^\gamma(q^2)  
\sigma^{\mu \nu} q_\nu \bigg] u_N
\end{equation}
where  $F_1^\gamma(q^2) $ and $F_2^\gamma(q^2)$ are the Dirac and Pauli
electromagnetic form factors, which are functions of the squared momentum
transfer. We will also use the Sachs form factors, which are linear 
combinations
of the Dirac and Pauli form factors
\begin{eqnarray}
G_E &= & F_1-\tau F_2 \nonumber \\
G_M &= & F_1 + F_2  
\end{eqnarray}
where $\tau \equiv -q^2/4M_N^2 >0$. 

The quark flavor structure of these form factors can be revealed by
writing the matrix elements of individual quark currents in terms of
form factors:
\begin{equation}
 <N|\bar q^j \gamma^\mu q^j   | N> \, \equiv  \, \bar u_N \bigg[ F_1^j
 (q^2) \gamma^\mu
+  {i \over {2M_N}}  F_2^j (q^2)  
\sigma^{\mu \nu} q_\nu \bigg] u_N \> 
\end{equation}
where $j = u,d,$ or $ s$; this 
defines the form factors $F_1^j$ and $F_2^j$. Then using 
definitions analogous to Eqn. (23), we can write 
\begin{eqnarray}
G_E^\gamma &=&  {2 \over 3} G_E^u
			- {1 \over 3} G_E^d
				 - {1 \over 3} G_E^s   \\
G_M^\gamma &=&  {2 \over 3} G_M^u
			- {1 \over 3} G_M^d
				 - {1 \over 3} G_M^s  \> .
\end{eqnarray}

In direct analogy to Eqn. (21), we have expressions for the neutral weak
form factors $G_E^Z$ and $G_M^Z$ in terms of the different quark flavor
components:
\begin{equation}
G_{E,M}^{Z} = (1 - {8 \over 3} \sin^2 \theta_W) G_{E,M}^u
 +(- 1 + {4 \over 3} \sin^2 \theta_W)  G_{E,M}^d
	+(- 1 + {4 \over 3} \sin^2 \theta_W)  G_{E,M}^s  \>.
\end{equation}
Again it is important to emphasize that the form factors 
$G_{E,M}^{u,d,s}$ appearing in
this expression are {\it exactly} the same as those in the electromagnetic
form factors in Eqns. (25,26).

Utilizing isospin symmetry,
one then can eliminate the up and down quark contributions to the
neutral weak form factors by using the proton
and neutron
electromagnetic form factors and obtain the expressions
\begin{equation}
G_{E,M}^{Z,p} = (1 - 4 \sin^2 \theta_W) G_{E,M}^{\gamma,p}
		- G_{E,M}^{\gamma,n} - G_{E,M}^{s} \>.
\end{equation}
This is a key result. It shows how the neutral weak form factors are related
to the electromagnetic form factors plus a contribution from
the strange (electric or magnetic) form factor. Thus measurement 
of the neutral weak form factor will allow (after combination with the
electromagnetic form factors) determination of the strange form factor
of interest.

The electromagnetic form factors present in Eqns. (22,23) are very
accurately known (1-2 \%) for the proton in the momentum transfer 
region $Q^2 < 1$ (GeV/c)${}^2$. The neutron form factors are not known
as accurately as the proton form factors (the electric form factor $G_E^n$
is at present rather poorly constrained by experiment), although 
considerable work
to improve our knowledge of these
quantities is in progress.
Thus, the present lack of knowledge of the 
neutron form factors will not significantly hinder the interpretation of
the neutral weak form factors.

In the derivation of  Eqn. (28), it was assumed that isospin symmetry
was exact. Electromagnetic and quark mass effects can cause small
violations of isospin symmetry and introduce corrections to this
relation.
The effects of isospin violation on the extraction of strange form
factors from neutral weak and electromagnetic form factors has been
treated in some detail in \cite{miller98}. In that work it is found
that these corrections are very small, generally less that about 1\%
of the electromagnetic form factors, and have only a minor effect
the extraction of the strange form factors. 

\begin{figure}
\centerline{\psfig{figure=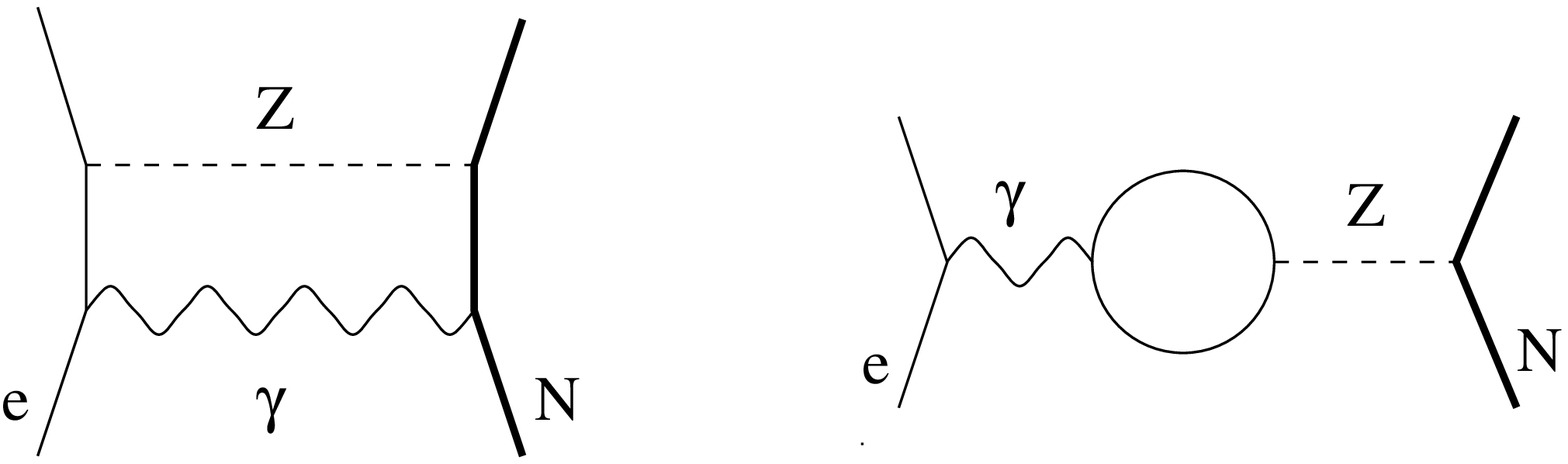,width=5.0in}}
\caption {Examples of amplitudes contributing to electroweak radiative
corrections (``$\gamma-Z$ box'' on the left) and anapole corrections
(``$\gamma-Z$ mixing'' on the right).}
\label{fig:radcorr}
\end{figure}

As mentioned in the previous section, there are electroweak radiative
corrections to the coefficients in Eqn. (28) due to processes such as
those shown in Figure~\ref{fig:radcorr}. The above expressions for the
neutral weak vector form factors $G^Z_{p,n}$ 
in terms of the electromagnetic form factors $G^\gamma_{p,n}$ 
are modified according to
\begin{equation}
G_{E,M}^{Z,p} = (1 -  4 \sin^2 \theta_W) 
(1+R^p_V) G_{E,M}^{\gamma,p}
- (1 + R^n_V) G_{E,M}^{\gamma,n} - G_{E,M}^{s} \>.
\end{equation}
The correction factors have been computed
 \cite{marciano, musolf90,musolf94a} 
to be 
\begin{eqnarray}
R^p_V &= &-0.054 \pm 0.033  \nonumber \\
R^n_V &= &-0.0143 \pm 0.0004.
\end{eqnarray}

\subsection{Strange Magnetism And The Strangeness Radius}

\begin{equation}
\mu_s \equiv G_M^s(Q^2=0)
\end{equation}
as the strange magnetic moment of the nucleon. Since the 
nucleon has no net strangeness, we find $G_E^s(Q^2=0) = 0$. However, 
one can express the slope of $G_E^s$ at $Q^2=0$
in the usual fashion in terms
of a ``strangeness radius'' $r_s$
\begin{equation}
r^2_s\equiv -6\left[dG_E^s/dQ^2\right]_{Q^2=0} \> .
\end{equation}

\subsection{Theoretical Models For The Strange Vector Form Factors}

\begin{figure}
\centerline{\psfig{figure=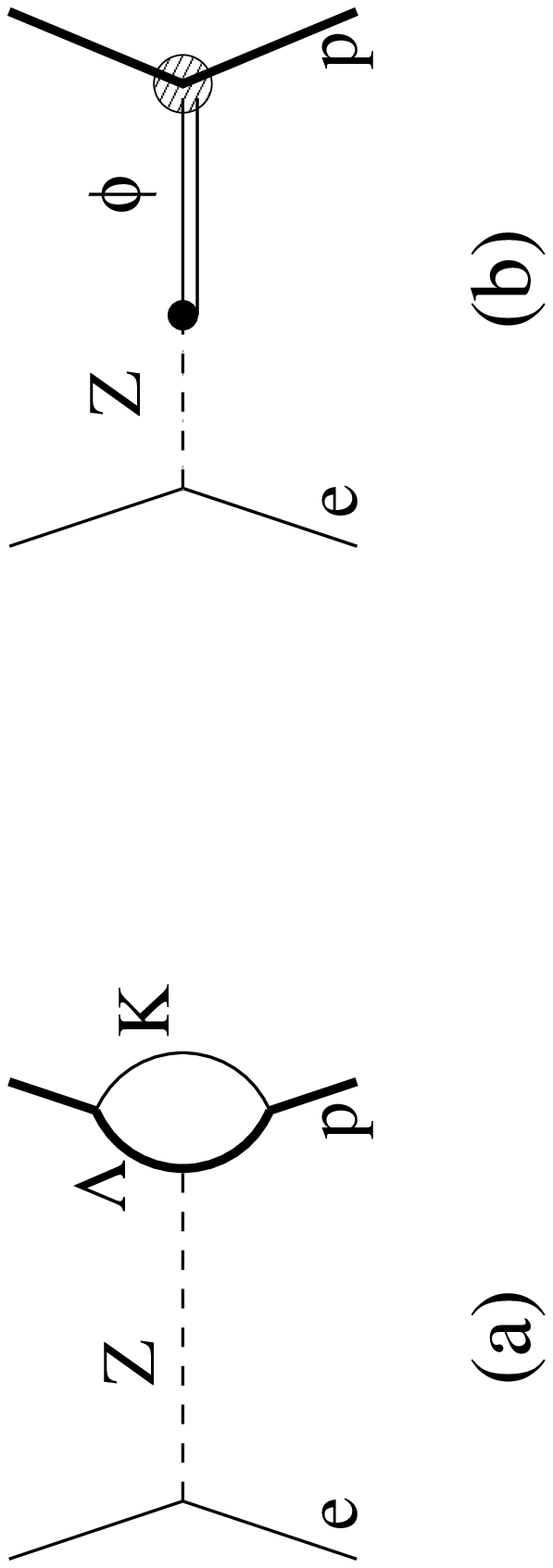,angle=270,width=5.0in}}
\caption {Examples of (a) loop and (b) pole diagrams used to compute
strangeness effects in the nucleon.}
\label{fig:loops}
\end{figure}

A variety of theoretical methods have been employed in efforts to
compute the form factors $G_{E,M}^s(Q^2)$ (or often just the
quantities $\mu_s$ and $r_s$).  Figure~\ref{fig:loops} shows two
examples of physical processes that may contribute. These are
generically known as ``loop'' effects and ``pole'' effects. The loop
effects~\cite{musolf94b,koepf,holstein,ito,geiger} correspond to the
fluctuation of the nucleon into a $K$-meson and hyperon. The physical
separation of the $s$ and $\bar s$ in such processes (or the
production of $s\bar s$ in a spin singlet) leads to non-zero values of
$G_{E,M}^s(Q^2)$.  The pole processes~\cite{jaffe89,hammer,meissner97}
are associated with the fluctuation of the virtual boson (photon
or $Z$) into a $\phi$ meson, which is predominantly a $\bar s s$ pair. Some
attempts have been made to combine the two approaches using dispersion
theoretical analyses~\cite{musolf96a}. Other models employ SU(3)
extensions of the Skyrme model~\cite{park,hong93,christov,hong97} or
the Nambu-Jona-Lasinio model.~\cite{weigel} Excited hyperons and
strange mesons are also included in some treatments, and these
contributions seem to be numerically significant.~\cite{geiger} A
detailed review of the various calculations can be found in
Ref.~\cite{beckHolstein}.

A reasonably complete compilation of theoretical
results for $\mu_s$ and $r_s^2$ is listed in Table~\ref{tab:calculations}. 
The calculated values of $r_s^2$ are small and there is no general
agreement on the sign. However, 
there is evidently a trend in Table~\ref{tab:calculations} that
one should expect $\mu_s <0$, generally in the range $-0.8 \to 0.0$ 
nuclear magnetons. Notable exceptions are references \cite{hong93}
and \cite{hong97} which analyze the set of baryon magnetic moments
in the context of a SU(3) generalization of the Skyrme model
Hamiltonian.

\begin{table}
\caption{Theoretical predictions for $\mu_s\equiv G_M^s(Q^2=0)$ and $r_s^2$.}
\label{tab:calculations}
\begin{tabular}{llll}
\hline
Type of calculation & $\mu_s$ (n.m.) & $r_s^2$(fm${}^2$) & Reference \\
\hline
Poles                   &$-0.31 \pm 0.09$               &$0.11\to 0.22$ & 
\cite{jaffe89}\\
Kaon Loops              &$-0.31 \rightarrow -0.40$      &$-0.032\to -0.027$&
 \cite{musolf94b} \\
Kaon Loops              &$-0.026$                       &-0.01&
\cite{koepf} \\
Kaon Loops              &$|\mu_s| = 0.8$                &       &
\cite{holstein}\\
SU(3) Skyrme (broken)   &$-0.13$                        &-0.10 &
\cite{park} \\
SU(3) Skyrme (symmetric) &$-0.33$                       &-0.19&
\cite{park} \\
SU(3) chiral hyperbag   &$+0.42$                        &       &
\cite{hong93} \\
SU(3) chiral color dielectric &$-0.20 \to -0.026$       &$-0.003\pm 0.002$&
 \cite{phatak} \\
SU(3) chiral soliton    &$-0.45 $               &-0.35&
\cite{christov} \\
Poles                   &$-0.24 \pm 0.03$               &$0.19 \pm 0.03$& 
\cite{hammer} \\
Kaon Loops              &$-0.125 \rightarrow -0.146$   &$-0.022\to -0.019$&
\cite{ito} \\
NJL soliton             &$-0.05 \rightarrow +0.25 $     &$-0.25\to -0.15$&
\cite{weigel} \\
QCD equalities          &$-0.75 \pm 0.30$               &       &
\cite{leinweber} \\
Loops                   &+0.035                 &-0.04  &
\cite{geiger} \\
Dispersion                   &$-0.10 \rightarrow -0.14$      &$0.21 \to0.27$&
\cite{musolf96a}\\
Chiral models           &$-0.25, -0.09$                 &0.24   &
\cite{musolf96b} \\
Poles                   &$0.003$                        &0.002&
\cite{meissner97}\\
SU(3) Skyrme (broken)   &$+0.36$                        &       &
\cite{hong97} \\
Lattice (quenched)   	&$-0.36 \pm 0.20$               & $ -0.06\to -0.16 $    &
\cite{dong98} \\
Lattice (chiral)   	&$-0.16 \pm 0.18$               &    &
\cite{leinweber00} \\

\hline
\end{tabular}
\end{table}

These model calculations unfortunately have a common difficulty -- the
need to, in one way or another, make important approximations to cover
unknown territory.  For example, the intuitively attractive picture of
a nucleon fluctuating into mesons and other baryons appears to have
very large contributions from higher mass states; in a simple (but
artificial) limit the contributions cancel.  Further, the dispersion
calculations have suggested that re-scattering is also important and
that the simple one-loop fluctuations are again just an approximation
and perhaps a rather poor one.  Heavy baryon chiral perturbation
theory, rather successful in other contexts, suffers generally from
the extension to three flavors in that apparently large and, in many
cases, unknown counterterms must be included to account for higher
order processes.  These difficulties, of course, stem from the fact
that experimentally we are separating out a ``microscopic'' quantity
-- the $\bar s s$ vector currents -- which we end up trying to
describe with ``macroscopic'' fields.  As is evident from
Table~\ref{tab:calculations} the present uncertainties are at the
100\% level.  There is hope that lattice calculations can help to sort
out these models utilizing effective degrees of freedom by looking at
quantities such as those associated with the sea quarks.  
Indeed, preliminary studies in the quenched approximation have already
been performed \cite{dong98,leinweber00}.
However, the first two-flavor unquenched calculations are now just
producing results and the extension to three flavors is non-trivial.
Accurate theoretical description of the strange quark vector currents
will therefore be an important topic in hadronic physics for some time
to come.

\section{AXIAL FORM FACTOR AND THE ANAPOLE MOMENT}

As noted above in Section 3, the parity-violating interaction of electrons
with nucleons involves an axial vector coupling to the nucleon, $G_A^e$.
This term in the parity-violating asymmetry contains several
effects beyond the leading
order $Z$- exchange which can only be differentiated in theoretical 
calculations.  Nevertheless,
it is important to establish that the
{\it experimentally observable} quantities are well-defined and unambiguous. 
To this end, we define the neutral weak axial form factors as observed in
neutrino scattering, $G_A^\nu$, and the corresponding quantity $G_A^e$, as 
indicated in the expression Eqn. (16). In the following, we discuss the 
relationship of each of these observables to nucleon beta decay, 
$W$- and $Z$-exchange, and higher order effects such as the anapole moment.

\subsection{Neutron Beta Decay And Elastic Neutrino Scattering}

The standard electroweak model relates the axial coupling, $G_A$, measured
in the charge-changing process (such as neutron beta decay) 
to the neutral process (such as neutrino scattering). For the case
of neutrino scattering, the situation is simplified due to the fact
that the neutrino has no (to lowest order) electromagnetic interaction.
However, due to the effect of $\bar s s $ pairs in generating the
isoscalar neutral weak form factor (see Table~\ref{tab:charges}), in lowest order 
we have the relation
\begin{equation}
G_A^\nu = G_A \tau_3 + \Delta s \> .
\label{eqn:gaz}
\end{equation}
$\Delta s $ is the same quantity that appeared in the discussion of
spin-dependent deep inelastic scattering in Section~\ref{sec:nucleonSpin} 
above.
$G_A (Q^2 =0) = 1.2601 \pm 0.0025$ is determined in neutron beta decay,
and the $Q^2$ dependence is measured in charged-current neutrino scattering
to be
\begin{equation}
G_A (Q^2) = {{G_A (Q^2 =0)} \over {(1+ {Q^2 \over M_A^2})^2 }}
\end{equation}
with $M_A \simeq 1.05$ GeV. In higher order, $G_A^\nu$ also contains
contributions from electroweak radiative corrections leading to the modified expression
\begin{equation}
G_A^\nu = G_A \tau_3 + \Delta s \>  + R_\nu
\end{equation}
where $R_\nu$ represents the radiative corrections which are expected to
be of order $\alpha$.

\subsection{Parity-Violating Electron Scattering: Higher Order Contributions}

In parity-violating electron scattering the 
neutral weak axial form factor corresponding
to tree-level $Z$-exchange is multiplied by the coefficient
$1 - 4 \sin^2 \theta_W \ll 1$. This suppression of the leading
amplitude increases the importance of anapole effects and other
electroweak radiative corrections:
\begin{equation}
G_A^e = G_A^Z + \eta F_A + R_e
\end{equation}
where 
\begin{equation}
\eta = {{8 \pi \sqrt{2} \alpha} \over {1 - 4 \sin^2 \theta_W}} = 3.45,
\end{equation}
$G_A^Z =  G_A \tau_3 + \Delta s$ (as in Eqn.~\ref{eqn:gaz}), 
$F_A$ is the nucleon anapole form factor (defined below), and 
$R_e$ are radiative
corrections. Typical contributions to $R_e$ and $F_A$ are shown in
Figure~\ref{fig:radcorr}.

As discussed in
\cite{musolf90,musolf94a}, the separation of $F_A$ and $R_e$ is actually a
theoretical issue and dependent upon the choice of gauge. 
In calculations performed to date  \cite{marciano,musolfphd}
the anapole type effects associated with the ``$\gamma - Z$ mixing'' 
amplitudes are, in fact, the dominant correction. We thus 
refer to the observable difference between $G_A^e$
and $G_A^\nu$ as an anapole contribution, with the caveat that 
the complete set of radiative corrections must be included in any 
consistent quantitative
theoretical treatment of $G_A^e$. 

The anapole moment has been traditionally defined as the effective
parity-violating
coupling between real photons and nucleons \cite{zeldovich}. 
(In practice, this quantity
is only observable at finite momentum transfer associated with the
parity-violating interaction between electrons and nucleons.) It appears
as an additional term in Eqn. (22) when one includes the possibility that
parity is not strictly conserved \cite{musolfphd}:
\begin{eqnarray}
\langle N| \hat V^\mu_\gamma  | N \rangle &\equiv& -e \, {\bar u_N} (p^\prime)
\{ F_1 \gamma^\mu - {i \over {2 m}} F_2 \sigma^{\mu \nu} q_\nu \nonumber \\
&+& F_A \, 
[ G_F (q^2 \gamma^\mu - q^\nu \gamma_\nu q^\mu ) \gamma^5 ]
\} u(p)
\end{eqnarray}
Note that our definition of $F_A$ differs from that used in the
atomic physics literature by a factor of $m^2 G_F$ with the
result that we expect the value of $F_A$ could be of order unity.
Thus, $F_A$ could indeed provide a substantial contribution to $G_A^e$ (see
Eqn. 36).

The anapole moment has been considered previously in atomic parity violation
experiments. Its definition is analogous to that in Eqn. 38 above, except that
it is now a form factor of the atomic nucleus (which may involve many nucleons).
In that case, it is expected that the anapole moment will be dominated by
many-body weak interaction effects in the nucleus \cite{haxton}. 
A value for the anapole moment of the Cesium atom has 
recently been reported \cite{weiman}.

We also note that parity-violating electron scattering
is not well suited to a determination of $\Delta s$ since this
contribution is suppressed and the corrections associated with
the nucleon axial vector couplings would obscure the interpretation.
It may be possible to achieve a better determination of $G_A^s$
in low energy neutrino scattering where the axial vector term
dominates the cross section and the radiative corrections are under better
control. (There is no suppression of the leading $Z$-exchange
amplitude and diagrams such as those in
Figure~\ref{fig:radcorr} involving a photon exchange do not
contribute to  neutrino scattering.)

\subsection{Theory Of The Anapole Contribution}

As mentioned above, 
aside from the leading $Z$ exchange
term ($G_A^Z$) the dominant calculated contribution to $G_A^e$
arises from the
``$\gamma - Z$ mixing'' diagram shown in Figure~\ref{fig:radcorr}
 \cite{marciano,musolfphd}.  It should be
noted that the evaluation of this amplitude
ignores the strong interaction of the nucleon with the
quark loop and so may not be numerically accurate. 
More recently, consideration of additional strong interaction
effects associated with mesonic processes
have indicated only relatively small additional corrections
 \cite{mus00,riska00,bira1,bira2}. The study of the anapole
contributions and other corrections
to $G_A^e$ is presently an active area of experimental and
theoretical investigation.

\section{THE EXPERIMENTAL PROGRAM}

\subsection{SAMPLE Experiment}
%{The SAMPLE Experiment}

The first experiment to determine the weak neutral
magnetic form factor of the proton is the SAMPLE experiment at the 
Bates Linear
Accelerator Center. Most of the modern techniques developed for the
measurement of small asymmetries in electron scattering experiments
are employed in this experiment. Thus some of the relevant methods
are described in some detail in this section. Nevertheless, only
a rather brief description is presented here; further
details are available in the Ph.D. theses of B. A. Mueller, 
T. A. Forest and D. P. Spayde~\cite{bryon,tony,damon}.

\begin{figure}
%\midinsert
\centerline{\psfig{figure=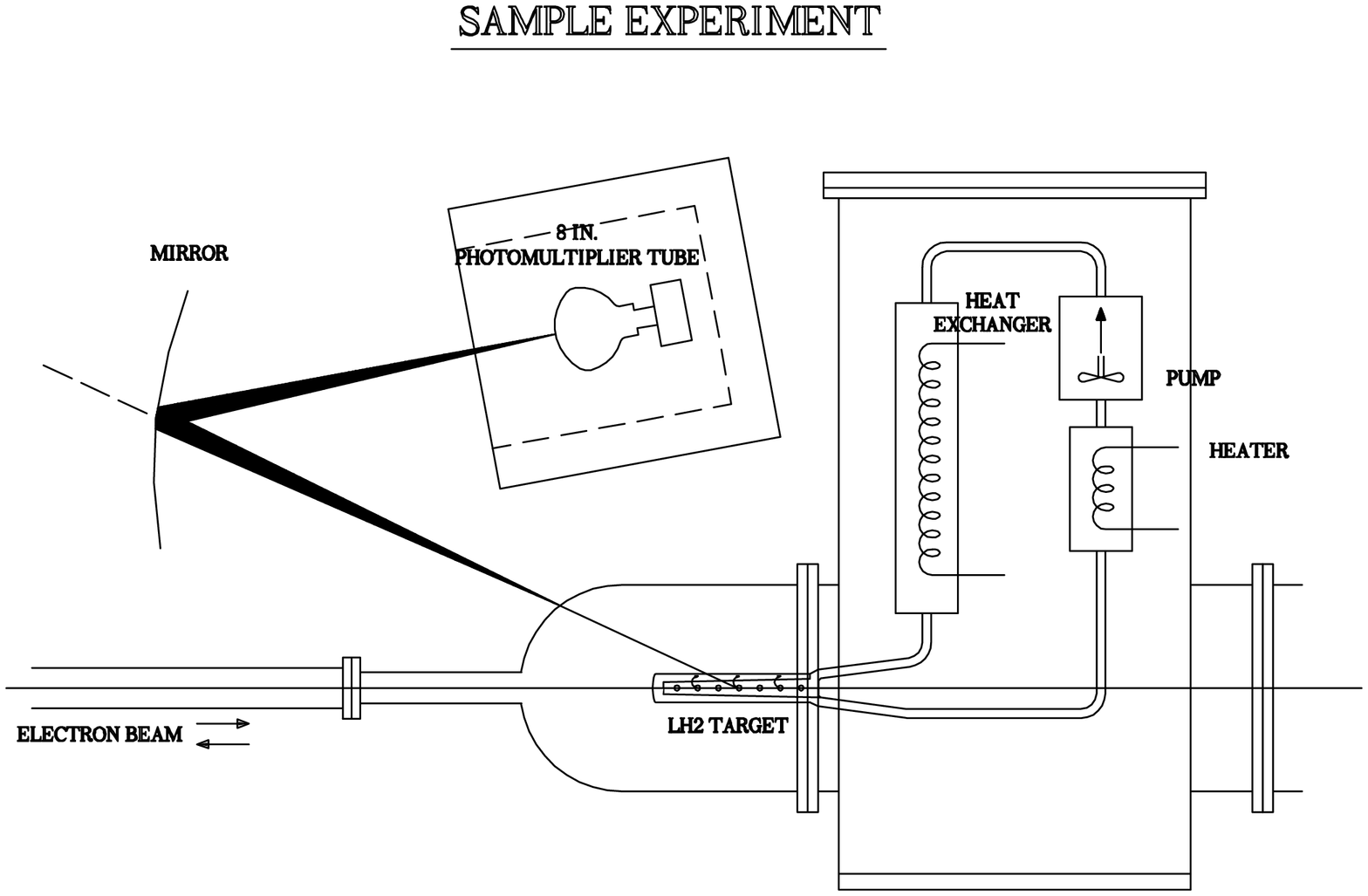,angle=90,height=3.0in}}
\caption {Schematic diagram of the layout of the SAMPLE target and
detector system.}
\label{fig:sampleSchematic}
%\caption{Schematic diagram of the layout of the SAMPLE target and
%detector system.}
\end{figure}

The experiment is performed using a 200 MeV polarized
electron beam incident on a liquid hydrogen target. The scattered electrons
are detected in a large solid angle ($\sim 1.5$ sr) air Cerenkov detector 
(similar to \cite{heil}) at backward
angles $130^\circ < \theta < 170^\circ$. This results in an 
average $Q^2 \simeq
0.1$ (GeV/$c)^2$. 
The expected asymmetry for $G_M^s=0$ is $-7.2 \times 10^{-6}$ or -7.2 ppm.
At these
kinematics the axial term is expected to contribute about 20\% 
of the asymmetry.
A schematic of the apparatus is shown in Figure~\ref{fig:sampleSchematic}.

The detector 
consists of 10 large ellipsoidal mirrors that reflect the Cerenkov
light into 8 inch diameter photomultiplier tubes. Each photomultiplier
is shielded from the target and room background by a cast lead shield.
The observed elastic scattering signal and background levels are generally
in accord with expectations.  The backgrounds involve 
non-Cerenkov sources of light which are diagnosed in separate measurements. 

Each detector signal is integrated over the
$\sim 25 \mu$sec of the beam pulse and digitized. The beam intensity
is similarly integrated and digitized. The ratio of integrated
detector signal to integrated
beam charge is the normalized yield which is proportional to cross section
(plus backgound).
The goal is to measure the beam helicity dependence of the cross section,
or equivalently, the helicity dependence of the normalized yield. All
10 detectors are combined in software during the data analysis.

The polarized electron source is a bulk GaAs photoemission source, with 
polarization typically 35\%.  The laser beam that is incident on the 
GaAs crystal is circularly polarized by a Pockels cell $\lambda /4$ plate.
The helicity is rapidly reversed by changing the voltage on the Pockels
cell to reverse the circular polarization of the light. 
An important capability is to manually reverse the beam helicity
by inserting a $\lambda$/2 plate which reverses the helicity of the light
(and the beam) relative to all electronic signals. A real parity violation
signal will change sign under this ``slow reversal''. Electronic
crosstalk and other effects will not change under ``slow reversal'', 
so this is an important test that the observed 
asymmetries are due to real physics rather than
spurious systematic effects in the experimental apparatus.

At the backward angles measured in the SAMPLE experiment,
the parity-violating asymmetry for elastic
scattering on the proton for the incident electron energy of 200~MeV
can be written as
\begin{eqnarray}
A_p &=& \left [ \frac{0.026}{\sigma_p} \right ]
\left [ \frac{-G_F Q^2}{\pi\alpha\sqrt{2}} \right ]
\left [ 1 - 0.27G^e_A(T=1) - 0.61G^s_M \right ] \nonumber \\
&=& -5.72 + 1.55G^e_A(T=1) + 3.49 G^s_M \> {\rm ppm}.
\end{eqnarray}
(The isoscalar component of $G_A^e$ is computed to be very small 
\cite{mus00} and we have
absorbed it into the leading constant term in these expressions.)
The reported measurement of this asymmetry \cite{spayde,mueller} by the 
SAMPLE collaboration is 
\begin{equation}
A_p = -4.92 \pm 0.61 \pm 0.73 \> {\rm ppm}.
\end{equation}

\begin{figure}
\centerline{\psfig{figure=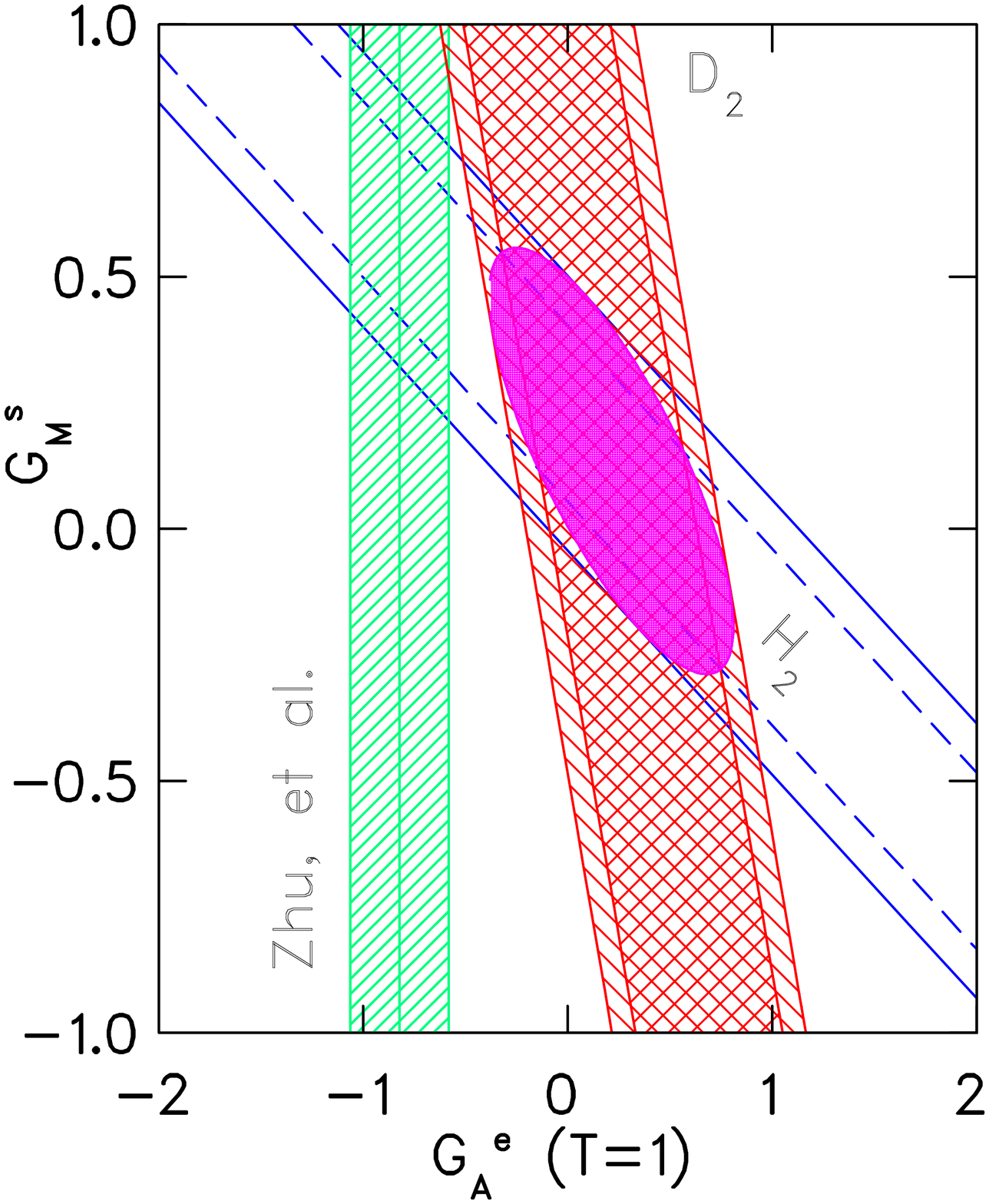,width=4.0in}}
%\centerline{\psfig{figure=sample_d2.ps,width=2.5in}}
%\vskip 0.1 in
\caption {Combined analysis of the data from the
two SAMPLE measurements. The two error bands from the hydrogen
experiment \cite{spayde} and the  deuterium
experiment \cite{hasty} are indicated. The inner hatched region includes the
statistical error and the outer represents the systematic uncertainty
added in quadrature.  Also plotted is the calculated
isovector axial $e$-$N$ form factor $G_A^e (T=1)$
obtained by using the anapole form factor and radiative 
corrections by Zhu {\it et al.} \cite{mus00}.}
\label{fig:sampleResult}
\end{figure}

The analogous expression for quasielastic electron scattering on the deuteron
at 200~MeV is
\begin{eqnarray}
A_d &=& \left [ \frac{0.049}{\sigma_d} \right ]
\left [ \frac{-G_F Q^2}{\pi\alpha\sqrt{2}} \right ]
\left [ 1 - 0.24G^e_A(T=1) - 0.10G^s_M \right ] \nonumber \\
&=& -7.27 + 1.78 G^e_A(T=1) + 0.75 G^s_M \> {\rm ppm}.
\end{eqnarray}
$\sigma_{(p,n)}$ are defined from the electromagnetic cross sections
$\sigma_{(p,n)}=\epsilon (G_E^{(p,n)})^2+\tau (G_M^{(p,n)})^2$,
where $\epsilon$ and $\tau$ are
kinematic factors. 
We have assumed the ``static approximation'' for the 
quasielastic scattering from the deuteron,
$\sigma_d =\sigma_p + \sigma_n$, as discussed and
studied in \cite{donnelly}. 
Thus we see that the deuteron asymmetry is relatively more sensitive to the
contribution from the isovector axial form factor $G^e_A(T=1)$
while the proton asymmetry has greater sensitivity to
the strange magnetic form factor.
Recently, the SAMPLE collaboration has reported
a value for this quasielastic deuterium asymmetry \cite{hasty}
\begin{equation}
A_d = -6.79 \pm 0.64 \pm 0.55 \> {\rm ppm} \> .
\end{equation}
The expected asymmetry using the tree level $G_A^e = G_A^Z$ and
$G_M^s=0$ is $A_d = -9.2$ ppm. 
%Need to evaluate correct value here!
Using the
calculations of $G_A^e$ in \cite{mus00} with $G_M^s=0$, the asymmetry 
is -8.7 ppm.  Thus the experimental result 
indicates that the substantial modifications of $G_A^e$
predicted in \cite{musolf90} are present, but probably with an even larger
magnitude than quoted in that work. 
It therefore appears that the neutral axial form factor determined
in electron scattering is substantially modified from the tree-level
$Z$-exchange amplitude (as determined in elastic $\nu$-$p$ scattering).
This is illustrated in Figure~\ref{fig:sampleResult}, 
where we show the allowed regions for
both the proton and deuteron SAMPLE results. This analysis indicates
that the magnetic strangeness is small
\begin{equation}
G_M^s (Q^2 = 0.1) = 0.14 \pm 0.29 \pm 0.31 \> .
\end{equation}
We can correct this value for the calculated $Q^2$ dependence \cite{hemmert}
of $G_M^s$ to obtain a result for the strange magnetic moment:
\begin{equation}
\mu_s = 0.01 \pm 0.29 \pm 0.31 \pm 0.07
\end{equation}
where the third uncertainty accounts for the additional uncertainty
associated with the theoretical extrapolation to $Q^2 = 0 $.

In addition, assuming the calculated small isoscalar axial corrections
are not grossly inaccurate, the isovector axial form factor can be
determined from the SAMPLE results
\begin{equation}
G_A^e(T=1) = +0.22 \pm 0.45 \pm 0.39
\end{equation}
in contrast with the calculated value \cite{mus00} $G_A^e (T=1) = -0.83 \pm
0.26$. This may be an indication that the anapole effects in the nucleon
are somewhat larger (by a factor of 2-3) 
than expected based on these calculations.

\subsection{The HAPPEX experiment}

The HAPPEX experiment \cite{hap,HAPPEX2} utilized the two
spectrometers in Hall A at Jefferson Lab to measure parity violation
in elastic electron scattering at very forward angles.  In this case,
the relatively small solid angle of each spectrometer, $\Delta\Omega =
5.5$ msr, is compensated by the very large (0.7 $\mu$b/sr ) 
cross section at forward angles
($\theta=12.3^\circ$) yielding, with a 15 cm long liquid hydrogen target,
a rate of roughly 1 MHz 
in each spectrometer.  The
scattered electrons were detected by integrating the output of a
simple lead-scintillator calorimeter.  This calorimeter was shaped to
accept only the elastic electrons, which are physically well separated
from the inelastic electrons in the focal plane of the spectrometer.  The
experiment was performed in two stages.  The first used a 100
$\mu$A beam with 39\% polarization produced from a bulk Ga-As crystal.
In the second, a strained Ga-As crystal was used, resulting in a beam
polarization of about 70\% and a current of 35 $\mu$A, slightly
improving the overall figure of merit (P$^2$I).

\begin{figure}[htbp]
\vspace*{13pt}
%\centerline{\vbox{\hrule width 5cm height0.001pt}}
\vspace*{2.81truein}		%ORIGINAL SIZE=1.6TRUEIN x 100% - 0.2TRUEIN
%\centerline{\vbox{\hrule width 5cm height0.001pt}}
\includegraphics{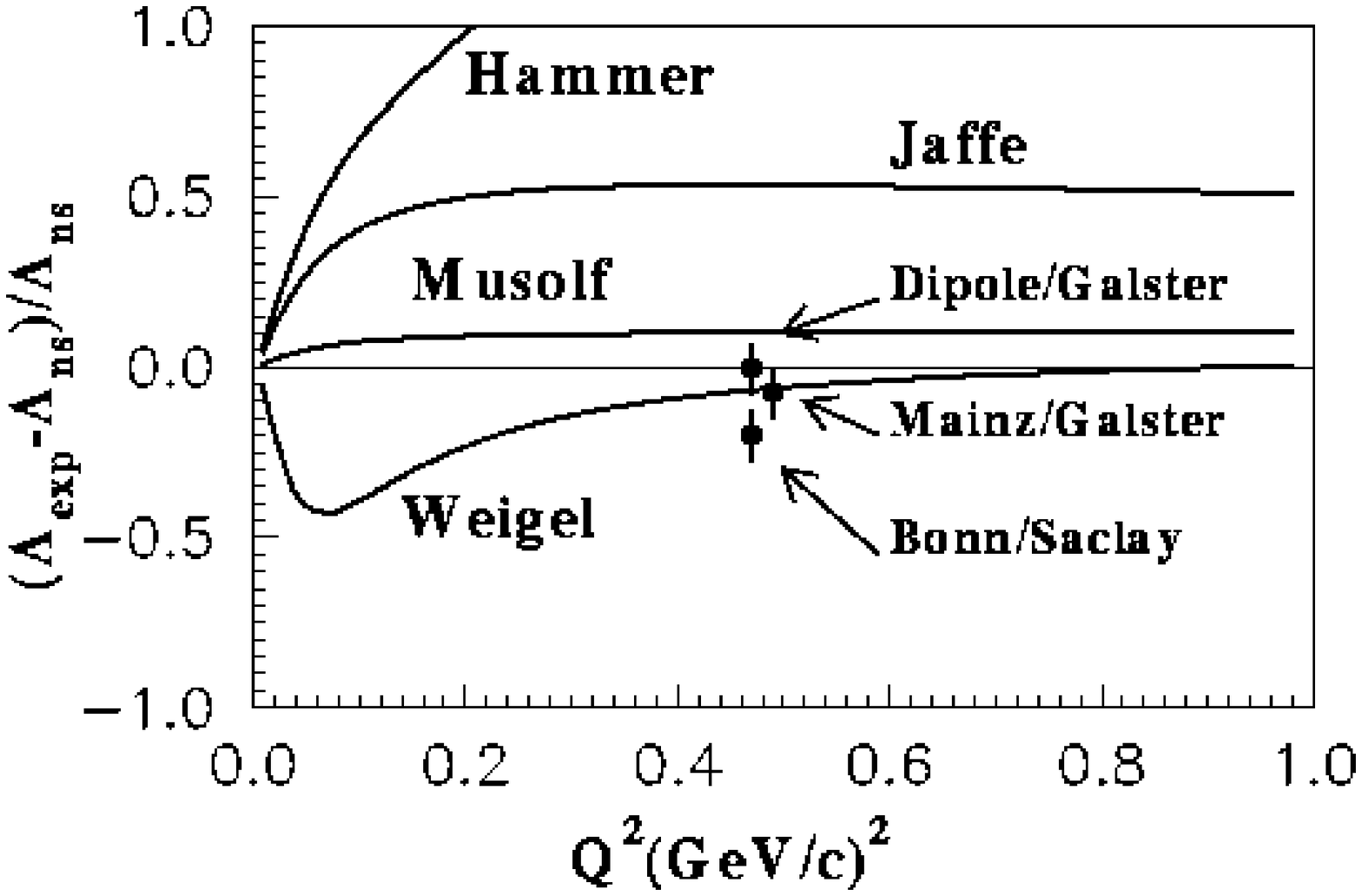}
\vspace*{13pt}
\caption{Results from the HAPPEX measurements of parity-violating electron 
scattering on hydrogen.  The three points on the plot correspond to
the neutron electromagnetic form factors of upper: $G_E^n$ from
Galster \cite{galster} and dipole $G_M^n$; middle $G_E^n$ from
Galster and $G_M^n$ from a Mainz
measurement \cite{mainzGMn}; and lower: $G_E^n$ from
a Saclay measurement \cite{platchkov} and $G_M^n$ from a Bonn
measurement \cite{Bonn}.  The calculations are from Hammer et
al. \cite{hammer}, Jaffe \cite{jaffe89}, Musolf and Ito \cite{musolf96b}, and Weigel,
et al. \cite{weigel}}
\label{fig:happexResult}
\end{figure}

The measured asymmetry, including the results from both phases of the
experiment is
\beqn
A_p(Q^2=0.477 \hbox{GeV}^2 \hbox{, }\theta_{av}=12.3^\circ) = -14.60
\pm 0.94 \pm 0.54~\hbox{ppm}
\eeqn
where again the first uncertainty is statistical and the second
systematic.  The largest sources of systematic uncertainty are
measurement of the beam polarization (3.2\% of its value) and
determination of $Q^2$ accruing from uncertainty in measurement of the
scattering (spectrometer) angle (contributing a 1.8\% uncertainty to
$A_p$).  The correction for beam induced false asymmetries amounted to only
$0.02 \pm 0.02$ ppm or about 0.1\% of $A_p$.
There is a significant uncertainty in the result owing from
uncertainty in the neutron form factors as is shown in
Fig.~\ref{fig:happexResult} and discussed further below.  It should be
noted that Fig.~\ref{fig:happexResult} suggests that the magnitude of
the measured asymmetry is less than that with no strange quarks, 
as is the case for the SAMPLE result.

Because the HAPPEX asymmetry was measured at a forward angle, it is in
principle sensitive to the three unmeasured form factors -- $G_E^s$,
$G_M^s$ and $G_A^e$.  The axial contribution is relatively small for
forward angles (becoming zero at $0^\circ$) and is expected to be $-0.56 \pm
0.23$ ppm out of the total of -14.6 ppm, assuming the calculated
value \cite{mus00} for $G_A^e(T=1)$ rather than that measured in the
SAMPLE experiment.  The other form factors enter in the combination
$G_E^s + 0.392 G_M^s$ for these kinematics.  The value of this
combination, normalized to the most accurately measured proton form
factor, $G_M^p/\mu_p$, is 
\beqn 
\frac{G_E^s + 0.392 G_M^s}{G_M^p/\mu_p}
= 0.091 \pm 0.054 \pm 0.039 
\eeqn 
where the first uncertainty is a
combination of the statistical and systematic uncertainties in the
asymmetry combined in quadrature with the above quoted
uncertainty in the axial
contribution, and the second is due to the uncertainty in the other
electromagnetic form factors~\cite{HAPPEX2}. The results are
particularly sensitive to the neutron magnetic form factor as can be
seen in Fig.~\ref{fig:happexResult}; using the
results from a different recent $G_M^n$ measurement \cite{Bonn} yields
\beqn \frac{G_E^s + 0.392 G_M^s}{G_M^p/\mu_p} = 0.146 \pm 0.054 \pm
0.047 \eeqn Several experiments are planned for the near 
future to reduce the uncertainty in $G_M^n$.  

\subsection {Future experiments}

A number of parity-violating electron scattering experiments are planned
for the near future.  In order to better understand the contributions of the
axial current, the SAMPLE experiment will continue with a second deuterium
measurement at a momentum transfer of 0.03 GeV$^2$.~\cite{tito}.  The HAPPEX group is also approved to make
a measurement \cite{HAPPEXII} at a momentum transfer of $Q^2=0.1$ GeV$^2$ utilizing
new septum magnets placed in front of the existing spectrometers.
These septa will allow measurements at more forward angles (roughly
$6^\circ$ scattering angle) in order to increase the cross section at
low momentum transfers and
hence the overall figure of merit.  Other recently approved or re-approved
parity violation measurements at JLab include one to determine the
neutron radius of the Pb nucleus \cite{Michaels} and a second to
measure the asymmetry in scattering from He at low momentum
transfer \cite{Armstrong} to measure the proton strangeness radius together 
with a He measurement at high momentum transfer \cite{Beise} 
($Q^2=0.6$ GeV$^2$) where early predictions showed a large value of $G_E^s$.
We note that because He is a $0^+$, $T=0$ nucleus, there are neither 
contributions from $G_M$ nor from $G_A$, making it particulary advantageous 
for measurements of $G_E^s$.

Two new parity violation experiments are being mounted with dedicated 
apparatus to address the
questions of the weak neutral current in the nucleon.  The PVA4
experiment~\cite{pva4}, underway at the MAMI accelerator in Mainz will
measure both forward and backward asymmetries using an array of
PbF$_2$ calorimeter crystals.  The G0 experiment~\cite{g0}, to be performed at
JLab, will also
measure at forward and backward angles to 
separate the contributions of the charge, magnetic and axial terms
over the full range of momentum transfers from about $Q^2=0.1$ to $Q^2=1.0$
GeV$^2$.

\begin{figure}[htbp]
\vspace*{13pt}
%\centerline{\vbox{\hrule width 5cm height0.001pt}}
\vspace*{3.0truein}		%ORIGINAL SIZE=1.6TRUEIN x 100% - 0.2TRUEIN
%\centerline{\vbox{\hrule width 5cm height0.001pt}}
%\centerline{\psfig{figure=Fig61s.ps,angle=270,height=3.0in}}
\includegraphics{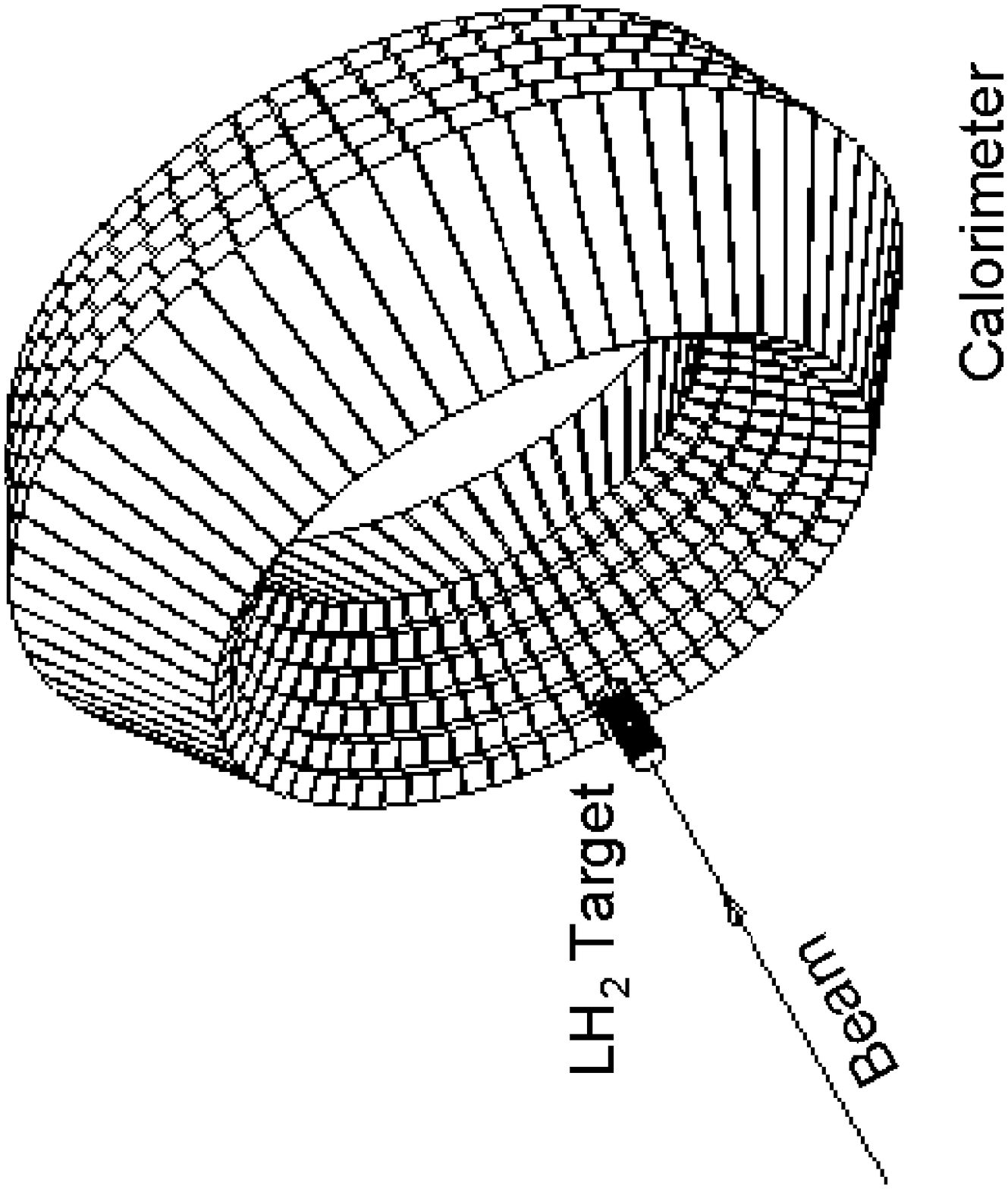}
\vspace*{13pt}
\caption{Schematic of the parity-violating electron scattering experimenta PVA4 
being performed at the Mainz Microtron.  The calorimeter consists of an array of 
1022 PbF$_2$ crystals used to count elastically scattered electrons.}
\label{fig:PVA4}
\end{figure}

The PVA4 experiment will initially utilize the 855 MeV beam from the
MAMI accelerator to measure the parity-violating elastic scattering
asymmetry at an angle centered around $35^\circ$ ($Q^2=0.23$ GeV$^2$).
This forward angle asymmetry will yield a measurement of the quantity
$G_E^s + 0.21 G_M^s$; the measurement began in summer
2000.  A 20 $\mu$A beam with 80\% polarization is incident on a
10 cm LH$_2$ target for the experiment.  The detector for the
experiment consists of 1022 PbF$_2$ calorimeter crystals covering
a solid angle of 0.7 sr and arranged in a pointing geometry relative
to the target as shown in Figure~\ref{fig:PVA4}.  The first
measurements will be made with half the detectors arranged in two
diametrically opposed quarters covering half the total azimuthal
angle.  The fast Cerenkov signal from the PbF$_2$ allows separation of
elastic and inelastic electrons in hardware.  Using an analog sum of
signals from a central detector and its eight nearest neighbors an
energy resolution of about 3.5\% has been achieved with an integration
gate of 20 ns.  This allows effective separation of elastic and
inelastic electrons -- the inelastic yield being about x10 larger than
that from elastic scattering.  The same apparatus can be reversed
relative to the beam to provide corresponding asymmetries over a range
of momentum transfers at backward angles (i.e. with a scattering
angle of $145^\circ$).

\begin{figure}[htbp]
\vspace*{13pt}
%\centerline{\vbox{\hrule width 5cm height0.001pt}}
\vspace*{3.0truein}		%ORIGINAL SIZE=1.6TRUEIN x 100% - 0.2TRUEIN
%\centerline{\vbox{\hrule width 5cm height0.001pt}}
%\centerline{\psfig{figure=schematic_bw.ps,height=3.0in}}
\includegraphics{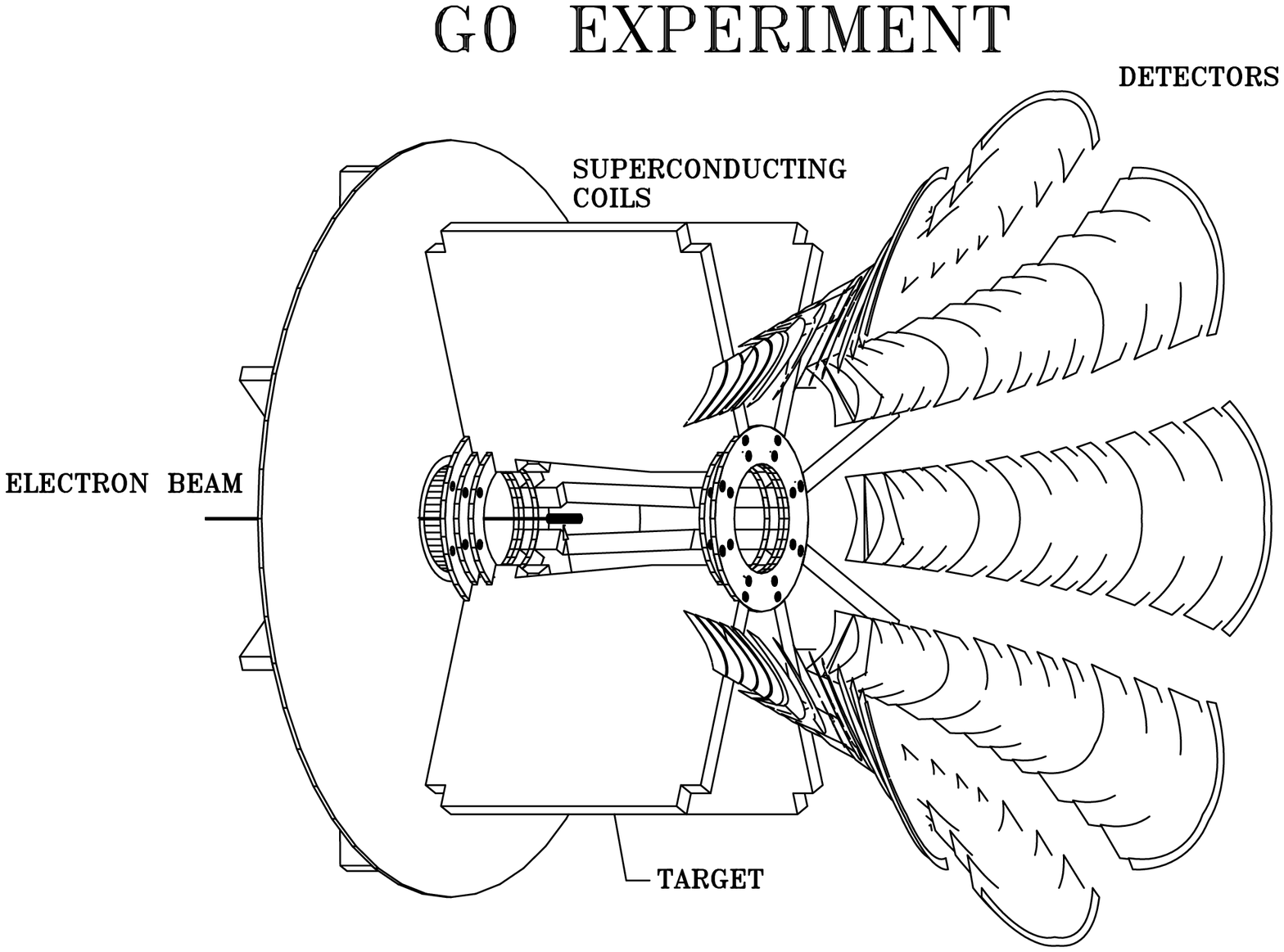}
\vspace*{13pt}
\caption{Schematic of the G0 parity-violating electron scattering experiment 
to 
be performed at JLab.  A dedicated superconducting toroidal 
spectrometer will be used to detect recoil protons for forward angle 
measurements and electrons for back angle measurements.}
\label{fig:G0}
\end{figure}

The goal of the G0 experiment is to measure forward proton asymmetries
and backward asymmetries for both the proton and deuteron in order to provide a
complete set of observables from which the charge, magnetic and axial
neutral weak currents of the nucleon can be determined.  It will
utilize a 40 $\mu$A, 70\% polarized beam from the JLab accelerator.
The experimental apparatus consists of a superconducting toroidal
magnet used to focus particles from a 20 cm liquid hydrogen target to
an array of plastic scintillator pairs located outside the magnet cryostat
(see Figure~\ref{fig:G0}).

In the first G0 experiment, forward angle asymmetries will be measured
by detecting the recoil protons from elastic scattering.  With an
acceptance of about 0.9 sr (for scattering angles centered at about
$70^\circ$), the spectrometer will measure asymmetries over the range
$0.12 < Q^2 < 1.0$ GeV$^2$ with a beam energy of 3 GeV.  For this
measurement elastic protons are identified by time-of-flight
(discriminating against inelastic protons and faster $\beta\sim 1$
particles such as $\pi^+$) and their $Q^2$ is determined by where in
the focal surface they are detected.  Custom electronics is being used
for fast accumulation of t.o.f. spectra with resolutions of 0.25 - 1
ns -- the maximum elastic rate in the scintillator pairs is about 1
MHz.  Background yields and asymmetries are thus measured concurrently
and will be used to correct the elastic asymmetries.

Backward angle asymmetries will be measured with the same apparatus,
by reversing it relative to the beam direction.  In this case
elastically scattered electrons will be measured at scattering angles
around $110^\circ$.  A set of smaller scintillators will be installed
near the exit window of the cryostat to discriminate elastic and
inelastic electrons.  In combination with the scintillators in the
focal surface this allows a rough measurement of both electron
momentum and scattering angle -- elastic electrons will appear only in
certain well defined pairs of detectors.  Measurements of
quasi-elastic scattering from deuterium at backward angles will
require improved particle identification
to separate electrons and $\pi^-$ (essentially
absent in the hydrogen measurements).  With this capability the G0 experiment
will be able to investigate the axial form factor $G_A^e$
over the full range of momentum transfers of the experiment.

\section{SUMMARY AND OUTLOOK}
%{Summary and Outlook}

As we have seen, the techniques developed to
measure parity-violating asymmetries in electron scattering 
can be utilized to address a variety of interesting and important
physics issues. 

In the immediate future, there is a program at
several laboratories to study the strange quark-antiquark contributions
to the vector form factors of the nucleon.
Although the presence of strange quarks and antiquarks in the nucleon
has been definitively established in deep inelastic neutrino
experiments, the role of these $\bar s s$ pairs in
determining the static properties of the nucleon is still, at present,
unclear. Analysis of the $\pi$-$N$ sigma term indicates that the
$\bar s s$ pairs may contribute $\sim 15\%$ to the mass of the
nucleon.
Future efforts to improve the experimental determination of the
$\pi$-$N$ amplitudes will be helpful, but it appears that the
theoretical
uncertainties associated with extracting the sigma term will remain.
It therefore appears unlikely that this situation will substantially
improve in the near future.
The contribution of strange quark-antiquark pairs to the nucleon spin
is also rather uncertain, again due to uncertainties in
theoretical interpretation. It appears that the best hope for
obtaining better information on $\Delta s$ lies in low-energy neutrino
scattering. However, technical progress in performing these difficult
experiments is required before reliable results can be obtained.
Nevertheless, this appears to be a promising avenue for development
of future experiments and the theoretical interpretation is on rather
solid ground.

Parity-violating electron scattering offers a more reliable method for 
determining the strange quark contributions to the charge and
magnetization
distributions of the nucleon. In principle, the experiments should
yield a detailed mapping of the $Q^2$ dependence (and therefore, the
spatial distribution) of the contributions from all three flavors of
quarks in the nucleon. We have seen the first glimpse of strangeness
in the nucleon's magnetization from the SAMPLE experiment,
along with evidence for an enhanced anapole effect in the axial form
factor $G_A^e$.

The SAMPLE and HAPPEX experiments will continue and several new experiments at
other laboratories are planned for the near future.  The PVA4 experiment at 
Mainz, already in progress, will provide a new forward angle asymmetry and 
possibly other back angle asymmetries as well.  The G0 experiment will provide
a complete separation of $G_E^s$, $G_M^s$ and $G_A^e$ over a momentum transfer
range of $0.1 - 1.0$ GeV$^2$.  Other measurements on $^4$He and $^{208}$Pb
will directly determine $G_E^s$ and the radius of the neutron distribution, 
respectively.

Clearly, the experimental program of studying parity violation in 
electron scattering presently offers great potential for providing
new and exciting information on fundamental interactions and 
hadronic structure. The experimental techniques continue to be
developed and improved, so we can expect new measurements with unprecedented
precision in the future. There is no question that this is a
subject with great promise - it only remains to be
seen how and when the various new secrets of nature will be revealed by
this experimental program.

\section{ACKNOWLEDGEMENTS}
%{Acknowledgments}

 This work was supported by
NSF grants PHY-0070182 and PHY-0072044.

\end{document}